\documentclass[a4paper,12pt]{article}
\advance\textheight2.7cm        
\advance\topmargin-2.0cm
\advance\textwidth2.6cm
\advance\evensidemargin-1.6cm   
\advance\oddsidemargin-1.6cm

\def\at#1{[*** \att #1 ***]}
\def\at#1{}           

\begin{document}


\begin{center}

{\LARGE \bf Phenomenological thermodynamics} \\

{\LARGE \bf in a nutshell} \\

\vspace{1cm}

\centerline{\sl {\large \bf Arnold Neumaier}}

\vspace{0.5cm}

\centerline{\sl Fakult\"at f\"ur Mathematik, Universit\"at Wien}
\centerline{\sl Oskar-Morgenstern-Platz 1, A-1090 Wien, Austria}
\centerline{\sl email: Arnold.Neumaier@univie.ac.at}
\centerline{\sl WWW: http://www.mat.univie.ac.at/\wave neum/}

\vspace{0.5cm}

April 21, 2014

\end{center}

\vspace{0.5cm}

\bfi{Abstract.} 
This paper gives a concise, mathematically rigorous description of 
phenomenological equilibrium thermodynamics for single-phase systems in 
the absence of chemical reactions and external forces. 

The present approach is similar to that of \sca{Callen} \cite{Cal}, who 
introduces in his well-known thermodynamics book the basic concepts by 
means of a few postulates from which everything else follows. 
His setting is modified to match the more fundamental approach based on 
statistical mechanics.  
Thermodynamic stability is derived from kinematical properties of 
states outside equilibrium by rigorous mathematical arguments,
superseding Callen's informal arguments that depend on a dynamical 
assumption close to equilibrium.

From the formulas provided, it is an easy step to go to various 
examples and applications discussed in standard textbooks such as
\sca{Callen} \cite{Cal} or \sca{Reichl} \cite{Rei}.
A full discussion of global equilibrium would also involve the
equilibrium treatment of multiple phases and chemical reactions. 
Since their discussion offers no new aspects compared with 
traditional textbook treatments, they are not treated here. \at{}

An older version of this document can be found as a chapter in the 
online book\\
\spd Arnold Neumaier and Dennis Westra,\\
\spd Classical and Quantum Mechanics via Lie algebras,\\
\spd 2011. \\
\spd http://lanl.arxiv.org/abs/0810.1019v2

\newpage
\tableofcontents 

\vspace*{1cm}

\section{Standard thermodynamical systems}\label{s.phen}

This paper gives a concise, mathematically rigorous description of 
phenomenological equilibrium thermodynamics for single-phase systems in 
the absence of chemical reactions and external forces. 
We discuss only the special but very important case of 
thermodynamic systems describing the single-phase global equilibrium of 
matter composed of one or several kinds of pure substances in the 
absence of chemical reactions and external forces. 
We call such systems \bfi{standard thermodynamic systems}; they are 
ubiquitous in the applications. 
In particular, a standard system is considered to be uncharged, 
homogeneous, and isotropic, so that each finite region looks like any 
other and is very large in microscopic units.

Pure substances of fixed chemical composition are labeled by an index 
$j$. Composite substances, called \bfi{mixtures}, are composed of
several pure substances called \bfi{components}. Each component has an 
index $j$; we denote by $J$ the set of all these indices.
A standard thermodynamic system is completely characterized 
by\footnote{
In the terminology, we mainly follow the IUPAC convention 
(\sca{Alberty} \cite[Section 7]{Alb}). 
For a history of thermodynamics notation, see \sca{Battino} et al. 
\cite{BatSW}.
} 
the \bfi{mole number} $n_j$ of each component $j$, the 
corresponding \bfi{chemical potential} $\mu_j$ of component $j$, the 
\bfi{volume} $V$, the \bfi{pressure} $P$, the \bfi{temperature} $T$, 
the \bfi{entropy} $S$, and the \bfi{internal energy} $U$. 
These variables, the 
\bfi{extensive} variables $n_j,V,S,U$ and the \bfi{intensive} variables 
$\mu_j,P,T$, are jointly called the \bfi{basic thermodynamic variables};
they take real numbers as values. (In this paper, all numbers are real.)
We group the $n_j$ and the $\mu_j$ into vectors $n$ and $\mu$ indexed 
by $J$ and write $\mu\cdot n = \D\sum_{j\in J} \mu_jn_j$. 
In the special case of a \bfi{pure substance}, there is just a single 
component; then we drop the indices and have $\mu\cdot n = \mu n$. 

Equilibrium thermodynamics is about characterizing so-called
equilibrium states in terms of intensive and extensive variables 
and their relations, and comparing them with similar nonequilibrium 
states. In a nonequilibrium state, only extensive variables have a 
well-defined meaning; but these are not sufficient to characterize 
system behavior completely.

All valid statements in the equilibrium thermodynamics of standard
systems can be deduced from the following definition. In this paper we 
take the properties asserted in the definition as axioms. However, they
can in turn be deduced from appropriate microscopic assumptions using
statistical \at{\footnote{This is the domain of statistical 
thermodynamics; see the companion paper.}}
mechanics; for example, convexity is a consequence of the 
statistical fact that covariance matrices are positive semidefinite. 

\begin{dfn}\bfi{(Phenomenological thermodynamics)}
\label{d.phen}\\
(i) Temperature $T$ and volume $V$ are positive, entropy $S$ and
mole numbers $n_j$ are nonnegative. 
The \bfi{extensive variables} $U,S,V,n_j$ are additive under the 
composition of disjoint subsystems. We combine the $n_j$ into a column 
vector with these components.

(ii) The \bfi{intensive variables} $T,P,\mu$ are related by the 
\bfi{equation of state}
\lbeq{e.def}
\Delta(T,P,\mu)=0.
\eeq
The \bfi{system function} $\Delta$ appearing in the equation of state 
is jointly convex \footnote{see Appendix \ref{s.convex}} in $T,P,\mu$ 
and decreasing in $P$. 
The set of $(T,P,\mu)$ satisfying $T>0$ and the equation of 
state is called the \bfi{state space}.\footnote{
Typically, the pressure $P$ is also positive. However, for solids, 
equilibrium states with $P<0$ are possible when exposed to tensile 
forces.
} 

(iii) The internal energy $U$ satisfies the \bfi{Euler inequality}
\lbeq{e.Ui}
U\ge T S - P V + \mu \cdot n
\eeq
for all $(T,P,\mu)$ in the state space.

(iv) \bfi{Equilibrium states} have well-defined intensive and extensive 
variables satisfying equality in \gzit{e.Ui}.
A system is in \bfi{equilibrium} if it is completely characterized by 
an equilibrium state.
\end{dfn}

This is the complete list of assumptions defining phenomenological 
equilibrium thermodynamics for standard systems; the system function 
$\Delta$ can be determined either by fitting to experimental data, 
or by calculation from a more fundamental description, namely the 
grand canonical partition function of statistical mechanics. 

All other properties follow from the system function.
Thus, all equilibrium properties of a material are characterized by 
the system function $\Delta$. (In contrast, nonequilibrium properties 
depend on additional dynamical assumptions, which, except close to 
equilibrium, vary among the various scientific schools. We will discuss
nonequilibrium properties only in passing.)

Surfaces where the system function is not differentiable 
correspond to so-called \bfi{phase transitions}. 
The equation of state shows that, apart from possible phase 
transitions, the state space has the 
structure of an $(s-1)$-dimensional manifold in $\Rz ^{s}$, where 
$s$ is the number of intensive variables; in case of a standard 
system, the manifold dimension is therefore one higher than the number 
of components.

Standard systems describe only a single phase of a substance 
(typically the solid, liquid, or gas phase), and changes between these 
as some thermodynamic variable(s) change.
Thermodynamic systems with multiple phases (e.g., boiling water, or 
water containing ice cubes) are only piecewise homogeneous. Each phase 
may be described separately as a standard thermodynamic system. 
But discussing the equilibrium at the interfaces between different 
phases needs some additional effort. (This is described in all 
common textbooks on thermodynamics.) \at{} Therefore, we consider only 
regions of the state space where the system function $\Delta$ is twice 
continuously differentiable.

\bigskip
Each equilibrium instance of the material is characterized by a
particular state $(T,P,\mu)$,
from which all equilibrium properties can be computed:

\begin{thm}~\\
(i) In any equilibrium state, the extensive variables are given by 
\lbeq{e.Sp}
S=\Omega\frac{\partial \Delta}{\partial T}(T,P,\mu),~~~
V=-\Omega\frac{\partial \Delta}{\partial P}(T,P,\mu),~~~
n=\Omega\frac{\partial \Delta}{\partial \mu}(T,P,\mu),
\eeq
and the \bfi{Euler equation}
\lbeq{e.Up}
U=T S - P V + \mu \cdot n,
\eeq
the case of equality in \gzit{e.Ui}.
Here $\Omega$ is a positive number independent of $T$, $P$, and $\mu$,
called the \bfi{system size}. 

(ii) In equilibrium, the matrix 
\lbeq{e.response}
\Sigma:=\left(\begin{array}{rrr}
\vspace{0.3cm}
\D\frac{\partial S}{\partial T}& 
\D\frac{\partial S}{\partial P}& 
\D\frac{\partial S}{\partial \mu} \\
\vspace{0.3cm}
\D-\frac{\partial V}{\partial T}& 
\D-\frac{\partial V}{\partial P}& 
\D-\frac{\partial V}{\partial \mu} \\
\D\frac{\partial n}{\partial T}& 
\D\frac{\partial n}{\partial P}& 
\D\frac{\partial n}{\partial\mu}
\end{array}\right)
\eeq
is symmetric and positive semidefinite. In particular, 
we have the \bfi{Maxwell reciprocity relations}
\lbeq{7-10p}
-\frac {\partial V} {\partial T}
=\frac {\partial S} {\partial P},~~~
\frac {\partial n_j} {\partial T}
=\frac {\partial S} {\partial\mu_j},~~~
\frac {\partial n_j} {\partial P}
=-\frac {\partial V} {\partial\mu_j},~~~
\frac {\partial n_j} {\partial\mu_k}
=\frac {\partial n_k} {\partial\mu_j}
\eeq
and the \bfi{stability conditions}
\lbeq{7-13p}
\frac {\partial S} {\partial T}\ge 0,~~~ 
\frac {\partial V} {\partial P}\le 0,~~~ 
\frac {\partial n_j} {\partial\mu_j}\ge 0.
\eeq
(iii) $\Delta(T,P,\mu)$ is monotone increasing in $T$ and each $\mu_i$, 
and strictly decreasing in $P$.
\end{thm}

\bepf
At fixed $S,V,n$, the inequality \gzit{e.Ui} holds in equilibrium with 
equality, by definition. Therefore the triple $(T,P,\mu)$ is a maximizer
of $TS-PV+\mu\cdot n$ under the constraints $\Delta(T,P,\mu)=0$, $T>0$. 
A necessary condition for a maximizer is the stationarity of 
the Lagrangian
\[
L(T,P,\mu)=TS-PV+\mu\cdot n -\Omega\Delta(T,P,\mu)
\]
for some Lagrange multiplier $\Omega$. Setting the partial derivatives 
to zero gives \gzit{e.Sp}, and since the maximum is attained in 
equilibrium, the Euler equation \gzit{e.Up} follows. The system size
$\Omega$ is positive since $V>0$ and $\Delta$ is decreasing in $P$.
The symmetry of the Hessian matrix
\[
\Delta''(T,P,\mu)=
\left(\begin{array}{rrr}
\vspace{0.3cm}
\D\frac{\partial^2\Delta}{\partial T^2}& 
\D\frac{\partial^2\Delta}{\partial P\partial T}& 
\D\frac{\partial^2\Delta}{\partial \mu\partial T} \\
\vspace{0.3cm}
\D\frac{\partial^2\Delta}{\partial T\partial P}& 
\D\frac{\partial^2\Delta}{\partial P^2}& 
\D\frac{\partial^2\Delta}{\partial \mu\partial P} \\
\D\frac{\partial^2\Delta}{\partial T \partial \mu}& 
\D\frac{\partial^2\Delta}{\partial P \partial \mu}& 
\D\frac{\partial^2\Delta}{\partial\mu^2}
\end{array}\right)
= \Omega^{-1}
\left(\begin{array}{rrr}
\vspace{0.3cm}
\D\frac{\partial S}{\partial T}& 
\D\frac{\partial S}{\partial P}& 
\D\frac{\partial S}{\partial \mu} \\
\vspace{0.3cm}
\D-\frac{\partial V}{\partial T}& 
\D-\frac{\partial V}{\partial P}& 
\D-\frac{\partial V}{\partial \mu} \\
\D\frac{\partial n}{\partial T}& 
\D\frac{\partial n}{\partial P}& 
\D\frac{\partial n}{\partial\mu}
\end{array}\right)
=\Omega^{-1}\Sigma
\]
of $\Delta$ implies the Maxwell reciprocity relations. Since 
$\Delta$ is convex, $\Sigma$ is positive semidefinite; hence the 
diagonal elements of $\Sigma$ are nonnegative, giving the stability 
conditions.
Finally, \gzit{e.Sp} implies that $\Delta$ is 
monotone increasing in $T$ and each $\mu_j$ since $S,n_j\ge 0$,
and strictly decreasing in $P$ since $V>0$. 
\epf

Many entries of the matrix $\Sigma$ from \gzit{e.response} are 
observable \bfi{response functions}. Note that not only the diagonal 
elements but the determinants of {\em all} principal submatrices of 
$\Sigma$ must be nonnegative. This gives further stability conditions.

Let $\mu_U$ and $\mu_S$ be constant vectors indexed by $J$.
Then replacing $\mu$, $U$, $S$, and $\Delta$ by 
\[
\mu':=\mu+\mu_U+T\mu_S,~~~U':=U+\mu_U\cdot n,~~~S':=S-\mu_S\cdot n,
\]
\[
\Delta'(T,P,\mu):=\Delta(T,P,\mu-\mu_U-T\mu_S)
\]
preserves \gzit{e.Sp} and \gzit{e.Up}, and hence all equilibrium 
properties. The existence of these \bfi{gauge transformations} implies 
that the chemical potentials are determined only up to a 
sub\-stance-dependent shift linear in the temperature, and the internal 
energy and entropy are only determined up to an arbitrary linear 
combinations of the mole numbers. This is an instance of the deeper 
problem to determine under which conditions thermodynamic variables 
are controllable.

\begin{expl}\label{ex.ideal}
The equilibrium behavior of electrically neutral gases at sufficiently 
low pressure can be modelled as ideal gases.
An \bfi{ideal gas} is defined by a system function of the form
\lbeq{e.ideal}
\Delta(T,P,\mu)=\sum_{j\in J}P_j(T) e^{\mu_j/ R T}-P,
\eeq
where the $P_j(T)$ are positive functions of the temperature,
\lbeq{e.R}
R\approx 8.31446\, JK^{-1}\mbox{mol}^{-1}
\eeq
is the \bfi{universal gas constant}\footnote{
For the internationally recommended values of this and other constants, 
their accuracy, determination, and history, see \sca{Mohr} et al. 
\cite{CODATA}.
}, 
and we use the bracketing convention $\mu_j/ R T =\mu_j/( R T)$.
Differentiation with respect to $P$ shows that $\Omega=V$ is the 
system size, and from \gzit{e.def}, \gzit{e.Sp}, and \gzit{e.Up}, we 
find that, in equilibrium,
\[
P=\sum_j P_j(T) e^{\mu_j/ R T},~~~
S=V\sum_j \Big(\frac{\partial}{\partial T}P_j(T) 
-\frac{\mu_jP_j(T)}{ R T^2}\Big) e^{\mu_j/ R T},
\]
\[
n_j =\frac{P_j(T)V}{ R T} e^{\mu_j/ R T},~~~
U=V\sum_j \Big(T\frac{\partial}{\partial T}P_j(T) 
-P_j(T)\Big) e^{\mu_j/ R T}.
\]
Expressed in terms of $T,V,n$, we have
\[
PV= R T \sum_j n_j,~~~\mu_j= R T\log\frac{ R T n_j}{P_j(T)V},~~~
U=\sum_j u_j(T)n_j,
\]
where, with $P-j'(T)=dP_j(T)/dT$, 
\lbeq{e.uj}
u_j(T)= R T\Big(\frac{TP_j'(T)}{P_j(T)} -1\Big),
\eeq
from which $S$ can be computed by means of the Euler equation 
\gzit{e.Up}. 
In particular, for one \bfi{mole} of a pure substance, defined by 
$n=1$, we get the \bfi{ideal gas law} 
\lbeq{e.ilaw}
PV=RT
\eeq
discovered by \sca{Clapeyron} \cite{Clap}; cf. \sca{Jensen} \cite{Jen}.

In general, the difference $u_j(T)-u_j(T')$ can be found 
experimentally by measuring the energy needed for raising or lowering 
the temperature of a component $j$ from $T'$ to $T$ while keeping 
the $n_j$ constant. In terms of infinitesimal increments, expressed 
through the \bfi{heat capacities} \at{define these more generally, 
as well as the other relevant thermodynamic quantities. Add the formula 
for heat capacities as sum of Schottky terms}
\[
C_j(T)=du_j(T)/dT,                             
\]
we have
\[
u_j(T)=u_j(T')+\int_{T'}^T dT\, C_j(T).
\]
From the definition \gzit{e.uj} of $u_j(T)$, we find that
\[
P_j(T)=P_j(T')\exp \int_{T'}^T \frac{dT}{T}
\Big(1+\frac{u_j(T)}{ R T}\Big).
\]
Thus there are two undetermined integration constants for each 
component. In view of the gauge transformations mentioned above, these 
can be chosen arbitrarily without changing the empirical content. 
The gauge transformation for $\Delta$ implies that $P_j(T)$ and 
$u_j(T)$ should be 
replaced by 
\[
P_j'(T):=e^{-(\mu_{Uj}+T\mu_{Sj})/ RT}P_j(T),~~~
u_j'(T):=u_j(T)+\mu_{Uj}.
\]
Therefore the partial pressures $P_j(T) e^{\mu_j/ R T}$ are 
unaffected by the gauge transformation and hence observable.
For an ideal gas, the gauge freedom can be fixed by choosing a 
particular \bfi{standard temperature} $T_0$ and setting arbitrarily 
$u_j(T_0)=0$, $\mu_j(T_0)=0$. 
Alternatively, at sufficiently large temperature $T$, heat capacities 
are usually nearly constant, and making use of the gauge freedom, 
we may simply assume that 
\[
u_j(T)=h_{j0} T,~~~P_j(T)=P_{j0} T \mbox{~~~for large~} T.
\]
\end{expl}

\section{The laws of thermodynamics}\label{s.cons}

In global equilibrium, all thermal variables are constant 
throughout the system, except at phase boundaries, where the extensive 
variables may exhibit jumps and only the intensive variables remain 
constant. This is sometimes referred to as the \bfi{zeroth law of 
thermodynamics}  (\sca{Fowler \& Guggenheim} \cite{FowG}) and 
characterizes global equilibrium; it allows one to measure intensive 
variables (like temperature) by bringing a calibrated instrument that 
is sensitive to this variable (for temperature a thermometer) into 
equilibrium with the system to be measured. 

For example, the ideal gas law \gzit{e.ilaw} can be used as a basis for 
the construction of a \bfi{gas thermometer}: The amount of expansion of 
volume in a long, thin tube can easily be read off from a scale along 
the tube. 
We have $V=aL$, where $a$ is the cross section area and $L$ is the 
length of the filled part of the tube, hence $T=(aP/R)L$. Thus, at 
constant pressure, the temperature of the gas is proportional to $L$. 
For the history of temperature, see 
\sca{Roller} \cite{Rol} and \sca{Truesdell} \cite{Tru}.

We say that two thermodynamic systems are brought in good \bfi{thermal 
contact} if, after a short time, the joint system tends to an 
equilibrium state. To measure the temperature of a system,
one brings it in thermal contact with a thermometer and waits until 
equilibrium is established.
The system and the thermometer will then have the same temperature, 
which can be read off from the thermometer. If the system is much larger
than the thermometer, this temperature will be essentially the same 
as the temperature of the system before the measurement.
For a survey of the problems involved in defining and
measuring temperature outside equilibrium, see 
\sca{Casas-V\'asquez \& Jou} \cite{CasJ}.

\bigskip
To be able to formulate the first law of thermodynamics we need the
concept of a reversible change of states, i.e., changes
preserving the equilibrium condition. For use in later sections,
we define the concept in a slightly more general form,
writing $\alpha$ for $P$ and $\mu$ jointly. We need to assume that the 
system under study is embedded into its environment in such a way that, 
at the boundary, certain thermodynamic variables are kept constant 
and independent of position. This determines the \bfi{boundary 
conditions} of the thermodynamic system; see the discussion in 
Section \ref{s.c1}.

\begin{dfn}
A \bfi{state variable} is an almost everywhere continuously 
differentiable function $\phi(T,\alpha)$ defined on the
state space (but sometimes only on an open and dense subset of it). 
Temporal changes in a state variable that 
occur when the boundary conditions are kept fixed are called 
\bfi{spontaneous changes}. 
A \bfi{reversible transformation} is a continuously differentiable 
mapping 
\[
\lambda \to (T(\lambda ),\alpha(\lambda ))
\]
from a real interval into the state space; thus
$\Delta(T(\lambda ),\alpha(\lambda ))=0$. 
The \bfi{differential}\footnote{
Informally, one may consider differentials as changes in states 
sufficiently small that a first order sensitivity analysis is 
appropriate: If $x$ changes by a small amount $dx$ then $y=f(x)$ 
changes by (approximately) the small amount $dy=f'(x)dx$, and 
analogous formulas hold in the multivariate case.\\ 
Another way to visualize equations or inequalities involving 
differentials is to view a reversible transformation as being 
realized (slowly but) continuously in time. 
Then the thermodynamic variables change continuously along a reversible 
path, and (as in the consideration of Carnot cycles) one can integrate 
the differentials along these paths, resulting in equations and 
inequalities of corresponding integrated quantities.\\
In formal mathematical terms, differentials are exact linear forms on 
the state space manifold; but we make no use of this. 
} 
\lbeq{3-12}
d\phi=\frac {\partial \phi} {\partial T}dT
+\frac {\partial \phi} {\partial \alpha} \cdot d\alpha, 
\eeq
obtained by multiplying the chain rule by $d\lambda$,
describes the change of a state variable $\phi$ under 
arbitrary (infinitesimal) reversible transformations. 
\end{dfn}

Reversible changes per se have nothing to do with changes in time. 
However, by sufficiently slow, quasistatic changes of the boundary 
conditions, reversible changes can often be realized approximately as 
temporal changes. The degree to which this is possible determines the 
efficiency of thermodynamic machines. The analysis of the efficiency
by means of the so-called \bfi{Carnot cycle} was the historical origin 
of thermodynamics.

The state space is often parameterized by 
different sets of state variables, as required by the application. 
If $T=T(\kappa,\lambda)$, $\alpha=\alpha(\kappa,\lambda)$ is such a
parameterization then the state variable $g(T,\alpha)$ can be written
as a function of $(\kappa,\lambda)$,
\lbeq{e.partial0}
g(\kappa,\lambda) = g(T(\kappa,\lambda),\alpha(\kappa,\lambda)).
\eeq
This notation, while mathematically ambiguous, is common in the
literature; the names of the argument decide which function is intended.
When writing partial derivatives without arguments, this leads to
serious ambiguities. These can be resolved by writing 
$\D\Big(\frac{\partial g}{\partial \lambda}\Big)_\kappa$ for the 
partial derivative of \gzit{e.partial0} with respect to $\lambda$;
it can be evaluated using \gzit{3-12}, giving the \bfi{chain rule}
\lbeq{e.partial}
\Big(\frac{\partial g}{\partial \lambda}\Big)_\kappa
=\frac{\partial g} {\partial T}
\Big(\frac{\partial T}{\partial \lambda}\Big)_\kappa
+\frac {\partial g} {\partial \alpha} \cdot
\Big(\frac{\partial \alpha}{\partial \lambda}\Big)_\kappa.
\eeq
Here the partial derivatives in the original 
parameterization by the intensive variables are written without 
parentheses.

Differentiating the equation of state \gzit{e.def}, using the chain 
rule \gzit{3-12}, and simplifying using \gzit{e.Sp} gives the 
\bfi{Gibbs-Duhem equation}
\lbeq{e.GDp}
0=SdT- VdP+n\cdot d\mu
\eeq
for reversible changes. If we differentiate the 
Euler equation \gzit{e.Up}, we obtain
\[
dU=TdS+SdT-PdV-VdP+\mu\cdot dn+n\cdot d\mu,
\]
and using \gzit{e.GDp}, this simplifies to the 
\bfi{first law of thermodynamics} 
\lbeq{3.1stp}
dU=TdS-Pd V +\mu \cdot dn.
\eeq
Historically, the first law of thermodynamics took on this form only 
gradually, through work by \sca{Mayer} \cite{May},
\sca{Joule} \cite{Jou}, \sca{Helmholtz} \cite{Hel}, and
\sca{Clausius} \cite{Cla}.

Considering global equilibrium from a fundamental point of view, the 
extensive variables are the variables that are conserved, or at least 
change so slowly that they may be regarded as time independent on the 
time scale of interest. In the absence of chemical reactions, the 
mole numbers, the entropy, and the internal energy are conserved; 
the volume is a system size variable which, in the fundamental view, 
must be taken as infinite (thermodynamic limit) to exclude the 
unavoidable interaction with the environment. 

However, real systems are always in contact with their environment,
and the conservation laws are approximate only. In thermodynamics, 
the description of the system boundary is generally reduced to the 
degrees of freedom observable at a given resolution.
The result of this reduced description (for derivations, see, e.g., 
\sca{Balian} \cite{Bal}, \sca{Grabert} \cite{Gra}, 
\sca{Rau \& M\"uller} \cite{RauM}) is a dynamical effect called 
\bfi{dissipation} (\sca{Thomson} \cite{Tho}). It is described by the 
\bfi{second law of thermodynamics}, which was discovered by
\sca{Clausius} \cite{Cla2}. In our context, we derive the second law as 
follows.

When viewing the Euler inequality \gzit{e.Ui} together with the Euler 
equation \gzit{e.Up}, parts (iii)-(iv) of Definition \ref{d.phen} say
that if $S,V,n$ are conserved (thermal, mechanical and chemical 
isolation) then the \bfi{internal energy},
\lbeq{e.int}
U:=TS-PV+\mu\cdot n
\eeq
is minimal in equilibrium.
\at{formulate more clearly; $T,P,\mu$ have no meaning outside
 equilibrium!}
If $T,V,n$ are conserved (mechanical and chemical isolation of a system 
at constant temperature $T$) then the \bfi{Helmholtz (free) energy},
\[
A:=U-TS=-PV+\mu\cdot n 
\]
is minimal in equilibrium.
If $T,P,n$ are conserved (chemical isolation of a system at constant 
temperature $T$ and pressure $P$) then the \bfi{Gibbs (free) energy},
\[
G:=A+PV=\mu\cdot n
\] 
is minimal in equilibrium.
If $S,P,n$ are conserved (thermal and chemical isolation of a system 
at constant pressure $P$) then the \bfi{enthalpy} 
\[
H:=U+PV=TS+\mu\cdot n
\] 
is minimal in equilibrium.
These rules just express the nondynamical part of the content of the 
second law since, in equilibrium thermodynamics, dynamical questions 
are ignored.

Finally, the \bfi{third law of thermodynamics}, due to \sca{Nernst} 
\cite{Ner}, says that the entropy is nonnegative. In view of 
\gzit{e.Sp}, this is equivalent to the monotonicity of 
$\Delta(T,P,\mu)$ with respect to $T$.

\section{Consequences of the first law}\label{s.c1}

The first law of thermodynamics describes the observable
energy balance in a reversible process. 
The total energy flux $dU$ into the system is composed of the 
\bfi{thermal energy flux} or \bfi{heat flux} $TdS$,
the \bfi{mechanical energy flux} $-PdV$, and the 
\bfi{chemical energy flux} $\mu \cdot dn$. 

The Gibbs-Duhem equation \gzit{e.GDp} describes the energy balance 
necessary to compensate the changes $d(TS)=TdS+SdT$ of
thermal energy, $d(PV)=Pd V + V dP$ of
mechanical energy, and $d(\mu \cdot n)=\mu \cdot dn+n\cdot d\mu$ 
of chemical energy in the energy contributions to the Euler equation 
to ensure that the Euler equation remains valid during a reversible 
transformation. Indeed, the Gibbs-Duhem equation and the first law
\gzit{3.1stp} together imply that $d(TS-PV+\mu\cdot n -U)$ vanishes, 
which expresses the invariance of the Euler equation (i.e., the 
equilibrium condition) under arbitrary reversible transformations. 

Related to the various energy fluxes are the \bfi{thermal work}
\[
Q = \int T(\lambda)dS(\lambda),
\]
the \bfi{mechanical work}
\[
W_\fns{mech} = -\int P(\lambda)dV(\lambda),
\]
and the \bfi{chemical work}
\[
W_\fns{chem} = \int \mu(\lambda)\cdot dn(\lambda)
\]
performed in a reversible transformation. The various kinds of work 
generally depend on the path through the state space; however, the
mechanical work depends only on the end points if the associated
process is conservative. 

As is apparent from the formulas given, thermal work is done by
changing the entropy of the system, mechanical work by changing the 
volume, and chemical work by changing the mole numbers. 
In particular, in case of thermal, mechanical, or chemical
\bfi{isolation}, the corresponding fluxes vanish identically. 
Thus, constant $S$ characterizes \bfi{adiabatic}, i.e., thermally 
isolated systems, constant $V$ characterizes mechanically 
isolated systems, and constant $n$ characterizes  
\bfi{closed}\footnote{
Note that the terms 'closed system' has also a 
much more general interpretation -- which we do {\em not} use in this 
chapter --, namely as a conservative dynamical system.
} 
(or \bfi{impermeable}) i.e., chemically isolated systems. 
Note that this constancy only holds when all assumptions for a 
standard system are valid: global equilibrium, a single phase, and 
the absence of chemical reactions.
Of course, these \bfi{boundary conditions} are somewhat idealized 
situations. But they can be approximately realized in practice and are 
of immense scientific and technological importance.

The first law shows that, in appropriate units, the temperature $T$ is 
the amount of internal energy needed to increase the entropy $S$ in a 
mechanically and chemically isolated system by one unit. 
The pressure $P$ is, in appropriate units, the amount of internal 
energy needed to decrease the volume $V$ in a thermally and chemically 
isolated system by one unit. In particular, increasing pressure 
decreases the volume; this explains the minus sign in the definition of 
$P$.
The chemical potential $\mu_j$ is, in appropriate units, the amount of 
internal energy needed to increase the mole number $n_j$ in a thermally 
and mechanically isolated system by one unit. 
With the traditional units, temperature, pressure, and chemical 
potentials are no longer energies.

We see that the entropy and the volume behave just like the mole
number. This analogy can be deepened by observing that mole numbers 
are the natural measure of the amounts of ``matter'' of each kind in a 
system, and chemical energy flux is accompanied by adding or removing 
matter.
Similarly, volume is the natural measure of the amount of ``space'' a
system occupies, and mechanical energy flux in a standard system is
accompanied by adding or removing space.
Thus we may regard entropy as the natural measure of the amount of 
``heat'' (colloquial) contained in a system\footnote{
Thus, entropy is the modern replacement for the historical concepts of 
\bfi{phlogiston} and \bfi{caloric}, which explained some heat phenomena
but failed to give a fully correct account of them. 
Phlogiston turned out to be ``missing oxygen'', an early 
analogue of the picture of positrons as``missing electrons'', holes
in the Dirac sea. Caloric was a massless substance of heat which had 
almost the right properties, explained many effects correctly, and 
fell out of favor only after it became known that caloric could be
generated in arbitrarily large amounts from mechanical energy, thus
discrediting the idea of heat being a substance. (For the precise 
relation of entropy and caloric, see \sca{Kuhn} \cite{Kuh1,Kuh2},
\sca{Walter} \cite{Walt}, and the references quoted there.) 
In the modern picture,
the extensivity of entropy models the substance-like properties of
the colloquial term ``heat''. But as there are no particles 
of space whose mole number is proportional to the volume, so there are 
no particles of heat whose mole number is proportional to the entropy.
Nevertheless, the introduction of heat particles on a formal level 
has some uses; see, e.g., \sca{Streater} \cite{Str}.
},  
since thermal energy flux is accompanied by adding or removing heat.
Looking at other extensive quantities, we also recognize energy as the 
natural measure of the amount of ``power'' (colloquial), 
momentum as the natural measure of the amount of ``force'' (colloquial),
and mass as the natural measure of the amount of ``inertia'' 
(colloquial) of a system. 
In each case, the notions in quotation marks are the colloquial terms 
which are associated in ordinary life with the more precise, formally 
defined physical quantities. For historical reasons, the words heat, 
power, and force are used in physics with a meaning different from the 
colloquial terms ``heat'', ``power'', and ``force''.

\section{Consequences of the second law}\label{s.c2}

The second law is centered around the impossibility of perpetual 
motion machines due to the inevitable loss of energy by  
dissipation such as friction (see, e.g., \sca{Bowden \& Leben} 
\cite{BowL}) or uncontrolled radiation.
This means that -- unless continually provided from the outside --
energy is lost with time until a metastable state is attained,
which usually is an equilibrium state. Therefore, the energy at
equilibrium is minimal under the circumstances dictated by the 
boundary conditions discussed before, which define the quantities held 
constant. 
In a purely kinematic setting as in our treatment, the approach to 
equilibrium cannot be studied, and only the minimal energy principles 
-- one for each set of boundary conditions -- remain. 

Traditionally, the second law is often expressed in the form of
an extremal principle for some thermodynamic potential.
For a complete list of thermodynamic potentials used in practice, see 
\sca{Alberty} \cite{Alb}. \at{add here and later enthalpy; note that 
heat of reaction is an enthalpy difference}
We derive here only the extremal principles for the internal energy,
the Helmholtz energy, and the Gibbs energy,\footnote{
The different potentials are related by so-called \bfi{Legendre 
transforms}; cf. \sca{Rockafellar} \cite{Roc} for the mathematical 
properties of Legendre transforms, \sca{Arnol'd} \cite{Arn} for their 
application in mechanics, and \sca{Alberty} \cite{Alb} for their 
application in chemistry.
} 
which give rise to the \bfi{internal energy potential}
\lbeq{e.Umax}
U(S,V,n) :=\max_{T,P,\mu}\,
\{TS-PV+\mu\cdot n\mid \Delta(T,P,\mu)=0;T>0\},
\eeq
the \bfi{Helmholtz potential}
\lbeq{e.Hmax}
A(T,V,n):=\max_{P,\mu}\,
\{-PV+\mu\cdot n\mid \Delta(T,P,\mu)=0\},
\eeq
and the \bfi{Gibbs potential}
\lbeq{e.Gmax}
G(T,P,n):=\max_\mu\,
\{\mu\cdot n\mid \Delta(T,P,\mu)=0\}.
\eeq
The arguments on the left hand side define the variables kept constant.

The Gibbs potential is of particular importance since everyday processes
frequently happen at approximately constant temperature, pressure, and 
mole number.

\begin{thm}\label{t.extstd}
\bfi{(Extremal principles)}\\
(i) In an arbitrary state, 
\lbeq{e.2ndU}
U \ge U(S,V,n),
\eeq
with equality iff the state is an equilibrium state. 
The remaining thermodynamic variables are then given by
\[
T = \frac{\partial}{\partial S}U(S,V,n),~~~
P = -\frac{\partial}{\partial V}U(S,V,n),~~~
\mu = \frac{\partial}{\partial n}U(S,V,n),~~~
U = U(S,V,n).
\]
In particular, an equilibrium state is uniquely determined by 
the values of $S$, $V$, and $n$.

(ii) In an arbitrary state, 
\lbeq{e.2ndA}
U-TS \ge A(T,V,n),
\eeq
with equality iff the state is an equilibrium state.
The remaining thermodynamic variables are then given by
\[
S=-\frac{\partial A}{\partial T}(T,V,n),~~~
P=-\frac{\partial A}{\partial V}(T,V,n),~~~
\mu=\frac{\partial A}{\partial n}(T,V,n),
\]
\[
U=A(T,V,n)+TS.
\]
In particular, an equilibrium state is uniquely determined by 
the values of $T$, $V$, and $n$.

(iii) In an arbitrary state,
\lbeq{e.2ndG}
U-TS+PV \ge G(T,P,n),
\eeq
with equality iff the state is an equilibrium state.
The remaining thermodynamic variables are then given by
\[
S=-\frac{\partial G}{\partial T}(T,P,n),~~~
V=\frac{\partial G}{\partial P}(T,P,n), ~~~
\mu=\frac{\partial G}{\partial n}(T,P,n),
\]
\[
U=G(T,P,n)+TS-PV.
\]
In particular, an equilibrium state is uniquely determined by 
the values of $T$, $P$, and $n$.
\end{thm}

\at{add enthalpy here and in next theorem}

\bepf
We prove (ii); the other two cases are entirely similar.
\gzit{e.2ndA} and the statement about equality are a direct consequence 
of Definition \ref{d.phen}(iii)--(iv). Thus, the difference 
$U-TS-A(T,V,n)$ takes its minimum value zero at the equilibrium value 
of $T$. 
Therefore, the derivative with respect to $T$ vanishes, which gives the
formula for $S$. To get the formulas for $P$ and $\mu$, we note that 
for constant $T$, the first law \gzit{3.1stp} implies
\[
dA=d(U-TS)=dU-TdS=-PdV+\mu\cdot dn.
\]
For a reversible transformation which only changes $P$ or $\mu_j$,
we conclude that $dA=-PdV$ and $dA=\mu_j dn_j$, respectively.
Solving for $P$ and $\mu_j$, respectively, implies the formulas for
$P$ and $\mu_j$.
\epf

The above results imply that one can regard each thermodynamic 
potential as a complete alternative way to describe the manifold of 
thermal states and hence all equilibrium properties.
This is very important in practice, where one usually describes
thermodynamic material properties in terms of the Helmholtz or Gibbs 
potential, using models like NRTL (\sca{Renon \& Prausnitz} \cite{RenP},
\sca{Prausnitz} et al. \cite{PraLA})
or SAFT (\sca{Chapman} et al. \cite{ChaGJR,ChaGJR2}).
The description in terms of the system function $\Delta$, although 
more fundamental from a theoretical perspective, is less useful since 
it expresses all quantities in terms of (temperature, pressure, and)
chemical potentials, and the latter are usually not directly accessible.
\at{but see the new paper! also phases make a difference!}

The additivity of extensive quantities is reflected in the 
corresponding properties of the thermodynamic potentials:

\begin{thm}\label{t.ext}
The potentials $U(S,V,n)$, $A(T,V,n)$, and $G(T,P,n)$
satisfy, for real $\lambda,\lambda^1,\lambda^2\ge 0$,
\lbeq{e.homUx}
U(\lambda S,\lambda V,\lambda n)=\lambda U(S,V,n),
\eeq
\lbeq{e.homAx}
A(T,\lambda V,\lambda n)=\lambda A(T,V,n),
\eeq
\lbeq{e.homGx}
G(T,P,\lambda n)=\lambda G(T,P,n),
\eeq
\lbeq{e.convUx}
U(\lambda^1 S^1+\lambda^2S^2,\lambda^1 V^1+\lambda^2V^2,
\lambda^1 n^1+\lambda^2n^2)
\le \lambda^1 U(S^1,V^1,n^1)+\lambda^2 U(S^2,V^2,n^2),
\eeq
\lbeq{e.convAx}
A(T,\lambda^1 V^1+\lambda^2V^2,\lambda^1 n^1+\lambda^2n^2)
\le \lambda^1 A(T,V^1,n^1)+\lambda^2 A(T,V^2,n^2),
\eeq
\lbeq{e.convGx}
G(T,P,\lambda^1 n^1+\lambda^2n^2)
\le \lambda^1 G(T,P,n^1)+\lambda^2 G(T,P,n^2).
\eeq
In particular, these potentials are convex in $S$, $V$, and $n$.
\end{thm}

\bepf
The first three equations express homogeneity and are a direct 
consequence of the definitions. Inequality \gzit{e.convAx} holds since, 
for suitable $P$ and $\mu$,
\[
\bary{lll}
A(T,\lambda^1 V^1+\lambda^2V^2,\lambda^1 n^1+\lambda^2n^2)
&=&-P(\lambda^1 V^1+\lambda^2V^2)+\mu\cdot(\lambda^1 n^1+\lambda^2n^2)\\
&=&\lambda^1(-PV^1+\mu\cdot n^1)+\lambda^2(-PV^2+\mu\cdot n^2)\\
&\le& \lambda^1 A(T,V^1,n^1)+\lambda^2 A(T,V^2,n^2);
\eary
\]
and the others follow in the same way.
Specialized to $\lambda^1+\lambda^2=1$, the inequalities express the
claimed convexity.
\epf

For a system at constant temperature $T$, pressure $P$, and mole
number $n$, consisting of
a number of parts labeled by a superscript $k$ which are separately
in equilibrium, the Gibbs energy is extensive (i.e., additive under 
composition of disjoint subsystems), since
\[
\bary{lll}
G&=&U-TS+PV= \D\sum U^k-T\sum S^k+P\sum V^k \\
&=& \D\sum (U^k-TS^k+PV^k)=\sum G^k.
\eary
\]
Equilibrium requires that $\sum G^k$ is minimal among all choices 
with $\sum n^k=n$, and by introducing a Lagrange multiplier vector 
$\mu^*$ for the constraints, we see that in equilibrium, the derivative 
of $\sum (G(T,P,n^k)-\mu^*\cdot n^k)$ with respect to each $n^k$ 
must vanish. This implies that 
\[
\mu^k= \frac{\partial G}{\partial n^k}(T,P,n^k)=\mu^*.
\]
Thus, in equilibrium, all $\mu^k$ must be the same.  
At constant $T$, $V$, and $n$, one can apply the same argument to the 
Helmholtz potential, at constant $S$, $V$, and $n$ to the 
internal energy potential, and at constant $S$, $P$, and $n$ to the 
enthalpy. In each case, the equilibrium is 
characterized by the constancy of the intensive parameters.

The second law may also be expressed in terms of entropy.

\begin{thm} \label{4.1.} (\bfi{Entropy form of the second law})\\
In an arbitrary state of a standard thermodynamic system 
\[
S \le S(U,V,n)
:=\min\,\{T^{-1}(U+PV-\mu\cdot n)\mid \Delta(T,P,\mu)=0\},
\]
with equality iff the state is an equilibrium state.
The remaining thermal variables are then given by
\lbeq{e.ent1x}
T^{-1}=\frac{\partial S}{\partial U}(U,V,n),~~~
T^{-1}P=\frac{\partial S}{\partial V}(U,V,n),~~~
T^{-1}\mu=-\frac{\partial S}{\partial n}(U,V,n),
\eeq
\lbeq{e.ent2x}
U=TS(T,V,n)-PV+\mu\cdot n.
\eeq
\end{thm}

\bepf
This is proved in the same way as Theorem \ref{t.extstd}.
\epf

This result -- perhaps the most famous but also most misunderstood 
version of the second law --
implies that when a system in which $U$, $V$ and $n$ are 
kept constant reaches equilibrium, the entropy must have increased.
Unfortunately, the assumption of constant $U$, $V$ and $n$
is unrealistic; such constraints are not easily realized in nature.
Under different constraints\footnote{
For example, if one pours milk into a cup of coffee, stirring mixes
coffee and milk, thus increasing complexity. Macroscopic order is
restored after some time when this increased complexity has become
macroscopically inaccessible. Since $T,P$ and $n$ are constant, 
the cup of coffee ends up in a 
state of minimal Gibbs energy, and not in a state of maximal entropy!
More formally, the first law shows that, for standard systems at fixed 
value of the mole number, the value of the entropy decreases 
when $U$ or $V$ (or both) decrease reversibly; this shows that the 
value of the entropy 
may well decrease if accompanied by a corresponding decrease of 
$U$ or $V$. The same holds out of equilibrium (though our 
equilibrium argument no longer applies).
For example, though it decreases the entropy, the reaction 
$2 H_2+O_2\to 2 H_2O$ may happen spontaneously at constant 
$T=25^\circ$C and $P=1$~atm if appropriately catalyzed.
}, 
the entropy is no longer maximal. 

The degree to which macroscopic space and time correlations are absent 
characterizes the amount of \bfi{macroscopic disorder} of a system.
Global equilibrium states are therefore macroscopically highly uniform; 
they are the most ordered macroscopic states in the universe rather 
than the most disordered ones. 
A system not in global equilibrium is characterized by macroscopic 
local inhomogeneities, indicating that the space-independent global 
equilibrium variables alone are not sufficient to describe the system.
Its intrinsic complexity is apparent only in a microscopic treatment.
The only macroscopic shadow of this complexity is the critical
opalescence of fluids near a critical point (\sca{Andrews} \cite{And}, 
\sca{Forster} \cite{For}). The contents of the second law of
thermodynamics for global equilibrium states may therefore be phrased 
informally as follows:
{\em In global equilibrium, macroscopic order (homogeneity) is perfect 
and microscopic complexity is maximal}.

In particular, the traditional interpretation of entropy as a measure of
disorder is often misleading.
Much more carefully argued support for this statement, with numerous
examples from teaching practice, is in \sca{Lambert} \cite{Lam}.
In systems with several phases, a naive interpretation of the second 
law as tendency moving systems towards increasing disorder is even more 
inappropriate: 
A mixture of water and oil spontaneously separates, thus ''ordering'' 
the water molecules and the oil molecules into separate phases! 

Thus, while the second law in the 
form of a maximum principle for the entropy has some theoretical and 
historical relevance, it is not the extremal principle ruling nature.
The irreversible nature of physical processes is instead manifest as 
\bfi{energy dissipation} which, in a microscopic interpretation, 
indicates the loss of energy to the unmodelled microscopic degrees of 
freedom.\footnote{
An example is friction, where macroscopic kinetic energy is translated 
into random motion of the molecules. Their details are not modelled
in a thermal description, and only their mean properties are reflected
-- via the so-called equipartition theorem -- in the temperature.
} 
Macroscopically, the global equilibrium states are therefore states 
of least free energy, the correct choice of which depends on the 
boundary condition, with the least possible freedom for change. 
This macroscopic immutability is another intuitive explanation for the 
maximal macroscopic order in global equilibrium states.

\section{The approach to equilibrium}\label{s.appeq}

Using only the present axioms, one can say a little bit about the
behavior of a system close to equilibrium in the following,
idealized situation.
Suppose that a system at constant $S$, $V$, and $n$ which is close to 
equilibrium at some time $t$ reaches equilibrium at some later time 
$t^*$. Then the second law implies 
\[
0\le U(t)-U(t^*) \approx (t-t^*)\frac{dU}{dt},
\]
so that $dU/dt\le 0$. We assume that the system is composed of two 
parts, which are both in equilibrium at times $t$ and $t^*$. Then
the time shift induces on both parts a reversible transformation, 
and the first law can be applied to them. Thus
\[
dU=\sum_{k=1,2} dU^k =\sum_{k=1,2} (T^kdS^k-P^kdV^k+\mu^k\cdot dn^k).
\]
Since $S$, $V$, and $n$ remain constant, we have $dS^1+dS^2=0$,
$dV^1+dV^2=0$, $dn^1+dn^2=0$, and since for the time shift $dU\le 0$,
we find the inequality
\[
0\ge (T^1-T^2)dS^1 - (P^1-P^2)dV^1 +(\mu^1-\mu^2)\cdot dn^1.
\]
This inequality gives information about the direction of the flow 
in case that all but one of the extensive variables are known to be 
fixed.

In particular, at constant $V^1$ and $n^1$, we have $dS^1\le 0$ if 
$T^1>T^2$; i.e., ``heat'' (entropy) flows from the hotter part towards 
the colder part. At constant $S^1$ and $n^1$, we have $dV^1\le 0$ if 
$P^1<P^2$; i.e., ``space'' (volume) flows from lower pressure to 
higher pressure: the volume of the lower pressure part decreases and 
is compensated by a corresponding increase of the volume in the higher 
pressure part. And for a pure substance at constant $S^1$ and 
$V^1$, we have $dn^1\le 0$ if $\mu^1>\mu^2$; i.e., ``matter'' (mole 
number) flows from higher chemical potential towards lower chemical 
potential. These qualitative results give temperature, pressure,
and chemical potential the familiar intuitive interpretation. 

This glimpse on nonequilibrium properties is a shadow of the far 
reaching fact that, in nonequilibrium 
thermodynamics, the intensive variables behave like potentials whose
gradients induce forces that tend to diminish these gradients,
thus enforcing (after the time needed to reach equilibrium) agreement 
of the intensive variables of different parts of a system. 
In particular, temperature acts as a thermal potential, whose 
differences create thermal forces which induce thermal currents, 
a flow of ``heat'' (entropy), in a similar way as differences in 
electrical potentials create electrical currents, a flow of 
``electricity'' (electrons)\footnote{
See \sca{Fuchs} \cite{Fuc} for a thermodynamics course (and for a 
German course \sca{Job} \cite{Job}) thoroughly exploiting these 
parallels. 
}. 
While these dynamical issues are outside the scope 
of the present work, they motivate the fact that one can control some
intensive parameters of the system by controlling the corresponding 
intensive parameters of the environment and making the walls permeable 
to the corresponding extensive quantities. This
corresponds to standard procedures familiar to everyone from ordinary 
life, such as heating to change the temperature, applying pressure 
to change the volume, or immersion into a substance to change the 
chemical composition.

The stronger nonequilibrium version of the second law says that 
(for suitable boundary conditions) equilibrium is actually attained 
after some time (strictly speaking, only in the limit of infinite time).
This implies that the energy difference 
\[
\delta E:=U-U(S,V,n)=U-TS-A(T,V,n)=U-TS+PV-G(T,P,n)
\]
is the amount of energy that is dissipated in order to reach 
equilibrium. In an equilibrium setting, we 
can only compare what happens to a system prepared in a nonequilibrium 
state assuming that, subsequently, the full energy difference 
$\delta E$ is dissipated so that the system ends up in an equilibrium 
state. Since few variables describe everything of interest, this 
constitutes the power of equilibrium thermodynamics. But this power is 
limited, since equilibrium thermodynamics is silent about when -- or 
whether at all -- equilibrium is reached. Indeed, in many cases, only 
metastable states are reached, which change too slowly to ever reach 
equilibrium on a human time scale. Typical examples of this are crystal 
defects, which constitute nonglobal minima of the free energy -- the 
global minimum would be a perfect crystal.

\appendix

\section{Convexity}\label{s.convex}

The mathematics of thermodynamics makes essential use of the concept 
of convexity. A set $X\subseteq \Rz^n$ is called \bfi{convex} if 
$tx+(1-t)y\in X$  for all $x,y\in X$ and all $t\in[0,1]$.
A real-valued function $\phi$ is called \bfi{convex} on the convex set
$X\subseteq \Rz^n$ if $\phi$ is defined on $X$ and, for all $x,y\in X$,
\[
\phi(tx+(1-t)y) \le t\phi(x)+(1-t)\phi(y) \for 0\le t\le 1.
\]
If $x$ is written explicitly as several arguments (such as $T,P,\mu$ 
in the main text), one says that $\phi$ is \bfi{jointly convex} in 
these arguments.
Clearly, $\phi$ is convex iff for all $x,y\in X$, the function 
$\mu:[0,1]\to \Rz$ defined by
\[
\mu(t):=\phi(x+t(y-x))
\]
is convex. It is well-known that, for twice continuously 
differentiable $\phi$, this is the case iff the second derivative 
$\mu''(t)$ is nonnegative for $0\le t\le 1$. 
Note that by a theorem of Aleksandrov (see \sca{Aleksandrov} \cite{Ale},
\sca{Alberti \& Ambrosio} \cite{AlbA}, \sca{Rockafellar} \cite{Roc.ca}),
convex functions are almost everywhere twice continuously 
differentiable: For almost every $x\in X$, there exist a 
unique vector $\partial\phi(x)\in \Rz^n$, the \bfi{gradient} of $\phi$ 
at $x$, and a unique symmetric, positive semidefinite matrix 
$\partial^2\phi(x)\in \Rz^{n\times n}$, the \bfi{Hessian} of $\phi$ 
at $x$, such that 
\[
\phi(x+h)=\phi(x)+h^T\partial\phi(x)
+\half h^T\partial^2\phi(x)h + o(\|h\|^2)
\]
for sufficiently small $h \in \Rz^n$.
A function $\phi$ is called \bfi{concave} if $-\phi$ is convex. 
Thus, for a twice continuously differentiable function $\phi$ of a 
single variable $\tau$, $\phi$ is concave iff $\mu''(\tau)\le 0$ for 
$0\le \tau\le 1$.

\addcontentsline{toc}{section}{References}
\bibliographystyle{plainurl}
\bibliography{/users/neum/ALL/books/physbook/QML1/QML.bib}


\end{document}